\documentclass[11pt]{article}

\newcommand{\etal}{{\it et al. }}

\usepackage{amssymb}

\usepackage[figuresright]{rotating}

\begin{document}





\begin{centering}
{\large\bf Hadronic final states at HERA\\[5mm]}


{P. J. Bussey\\for the H1 and ZEUS Collaborations\\[3mm]
School of Physics and Astronomy, University of Glasgow, Glasgow \mbox{G12 8QQ,} United Kingdom\\[3mm]
Talk presented at the XXXVII International Conference on High Energy Physics, Valencia, Spain, 2-9 July, 2014}

\begin{abstract}
Recent measurements of hadronic final states by H1 and ZEUS at HERA are presented.
The H1 measurements  consist of measurements of charged particle spectra in deep-inelastic $ep$ scattering
and of forward photons and neutrons.  The ZEUS results consist of a series of measurements 
of prompt photons in photoproduction.
\end{abstract}

\end{centering}





\section{Introduction}
This talk presents some recent measurements of hadronic final states
from the H1 and ZEUS Collaborations at HERA.  From H1, measurements of charged
particle spectra in deep-inelastic scattering (DIS) and of forward
photons and neutrons are shown.  The ZEUS results consist of a series
of measurements of prompt photons in photoproduction. More complete
details of the theoretical models that are used can be found in the
referenced papers and the citations therein.

\section{Charged-particle spectra in deep-inelastic scattering}
The H1 Collaboration have measured the distributions of charged
particles in $ep$ DIS, and compared the results with theoretical
models~\cite{h11}.  The usual variables are used, namely the virtual
photon virtuality $Q^2$ and $y$ defined as the fractional energy loss
of the lepton in the proton rest frame.  The variable $x$ is defined
as $Q^2/sy$, and the variables are defined in the hadronic
centre-of-mass frame.  In this analysis (only), the $+Z^*$ axis is
taken in the direction of the virtual photon.  Charged-particle
densities are integrated over the range $5<Q^2<100$ GeV$^2$.

As $Q^2$ and $x$ decrease, the evolution of the scattering
process should change from a DGLAP to a BFKL mechanism.  These form the basis of the
theoretical models tested, and there is a further model, CCFM, which is a
combination of these two approaches.  The models differ in details
concerning the transverse-momentum $p_T$ ordering of the radiated
partons in the calculations of the process.  Also, RAPGAP uses DGLAP
evolution, while DJANGOH uses a colour dipole model and a BFKL-like evolution,
and CASCADE uses CCFM.  The HERWIG model uses the POWHEG option.

The results show that none of the models
tested agrees with the data very well over the entire measured $p_T^*$
range.  However DJANGOH does best.  Further cross sections (not
presented here) in bins of $Q^2$ and $x$ show that at low $p_T^*$ the
distributions in $\eta^*$ are satisfactorily described by all the
models except CASCADE while at higher $p_T^*$ values, none of
the models is satisfactory except for DJANGOH.

\begin{figure}[t!]
~\\[-25mm]
\begin{centering}
\includegraphics[width=80mm]{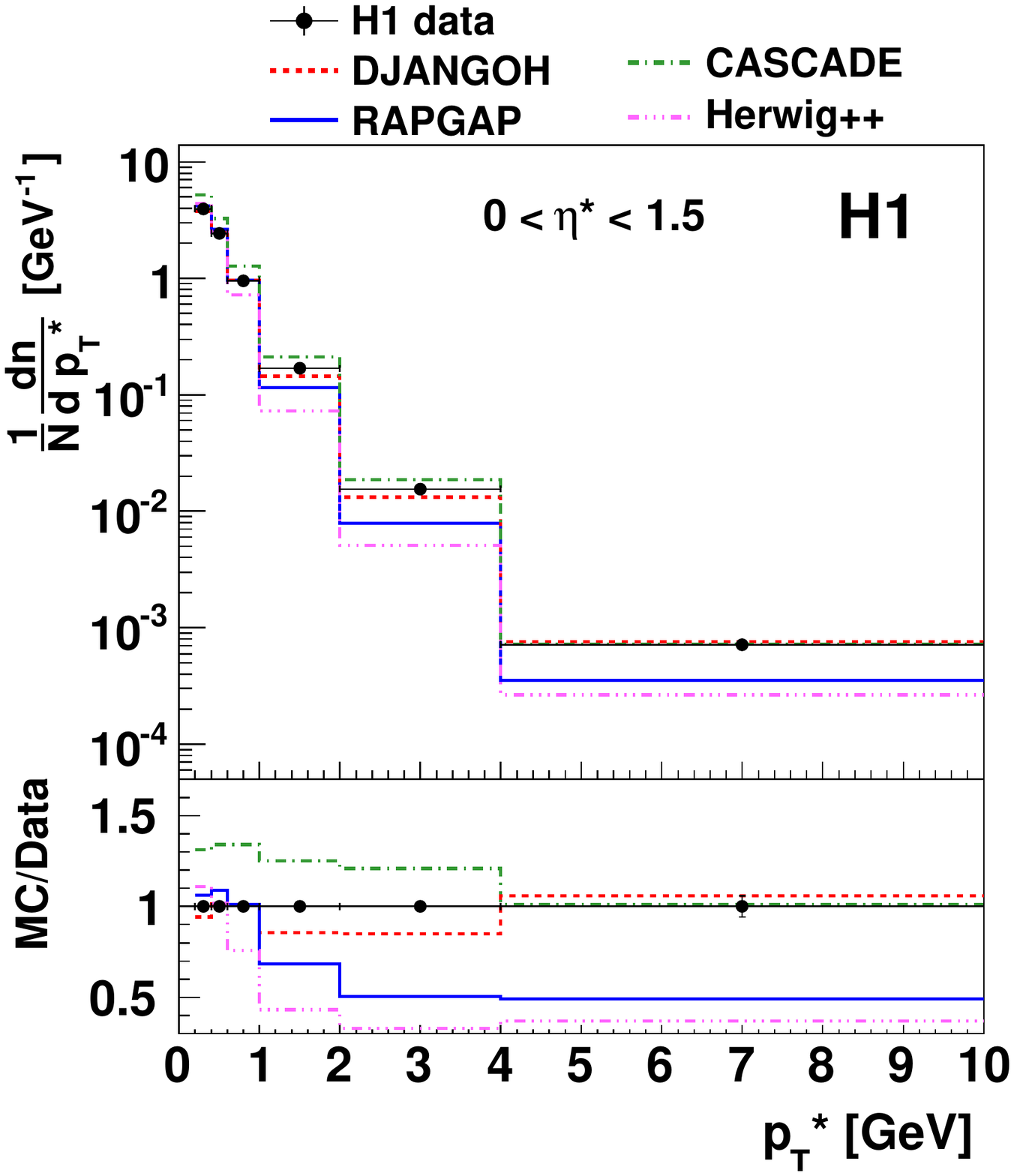}\\[-55.3mm]
\hspace*{26mm}(a)\\[22mm]
\includegraphics[width=80mm]{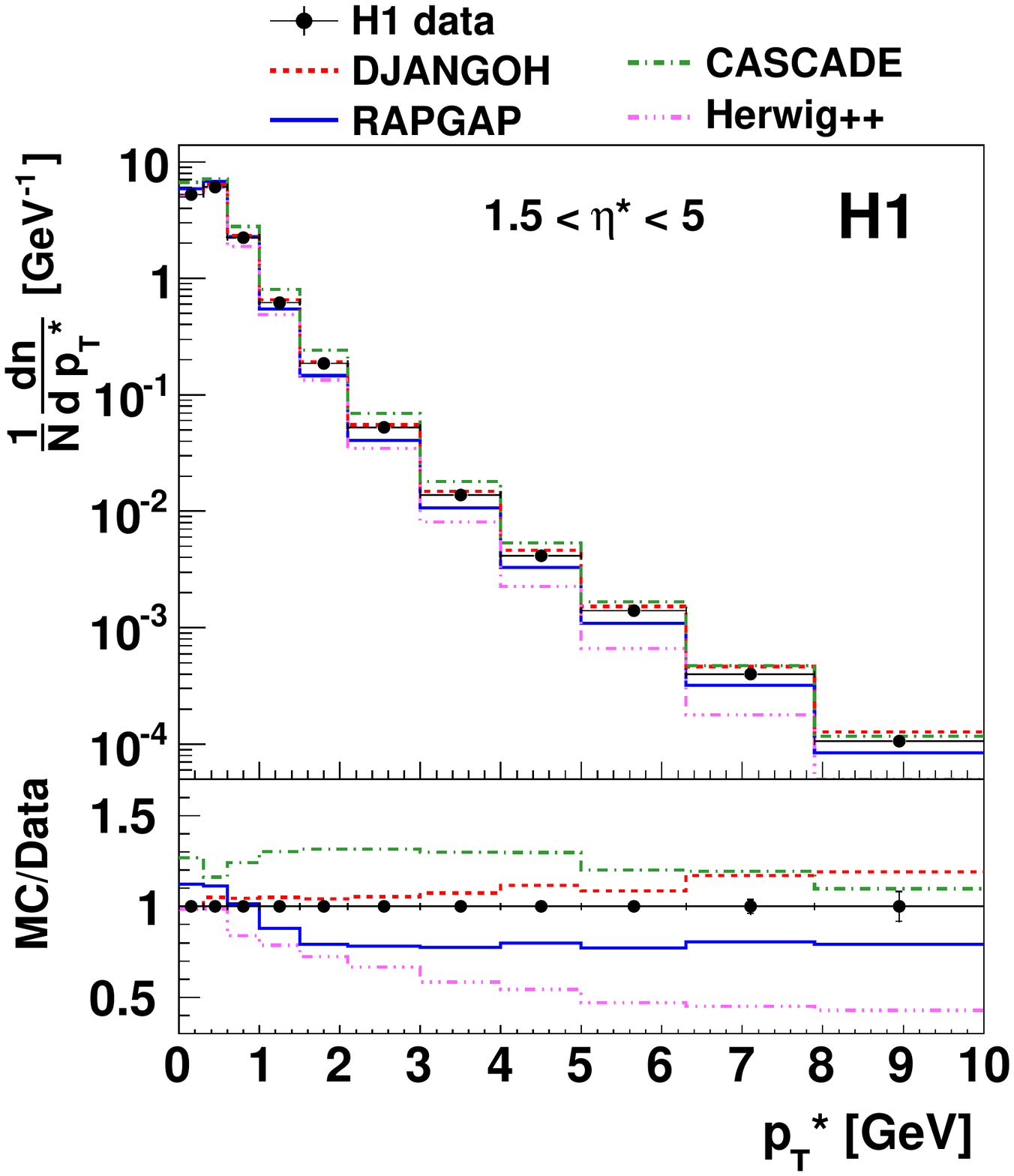}\\[-55.3mm]
\hspace*{26mm}(b)\\[41.4mm]
\end{centering}
\caption{Normalised particle densities from H1: (a) for 
central pseudorapidities, (b) for forward pseudorapidities. }
\label{fig1}  

\end{figure}

\begin{figure}[h!]
~\\[-12mm]
\begin{centering}
\mbox{\hspace*{-1.5mm}
\includegraphics[width=40mm]{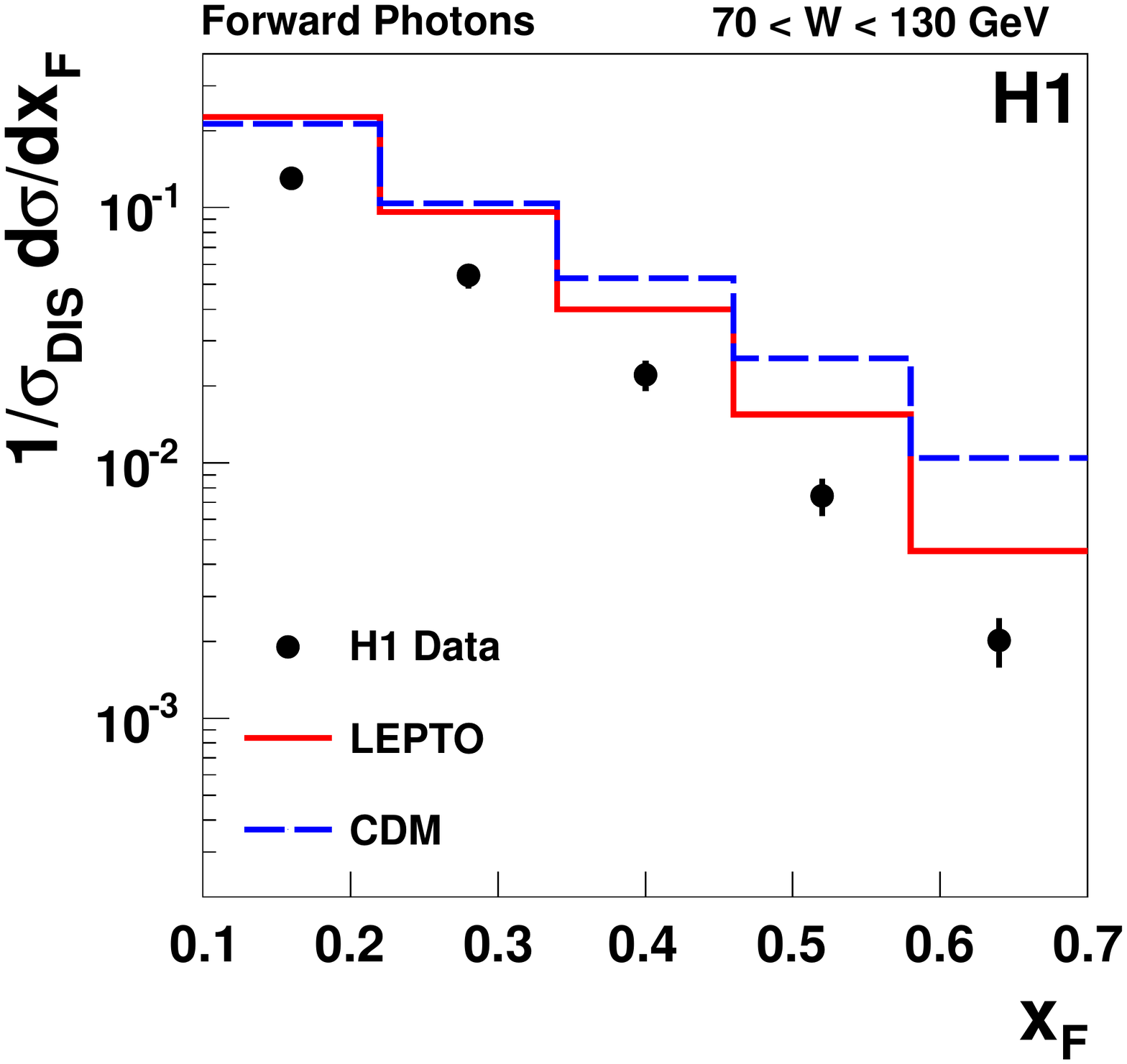}
\includegraphics[width=40mm]{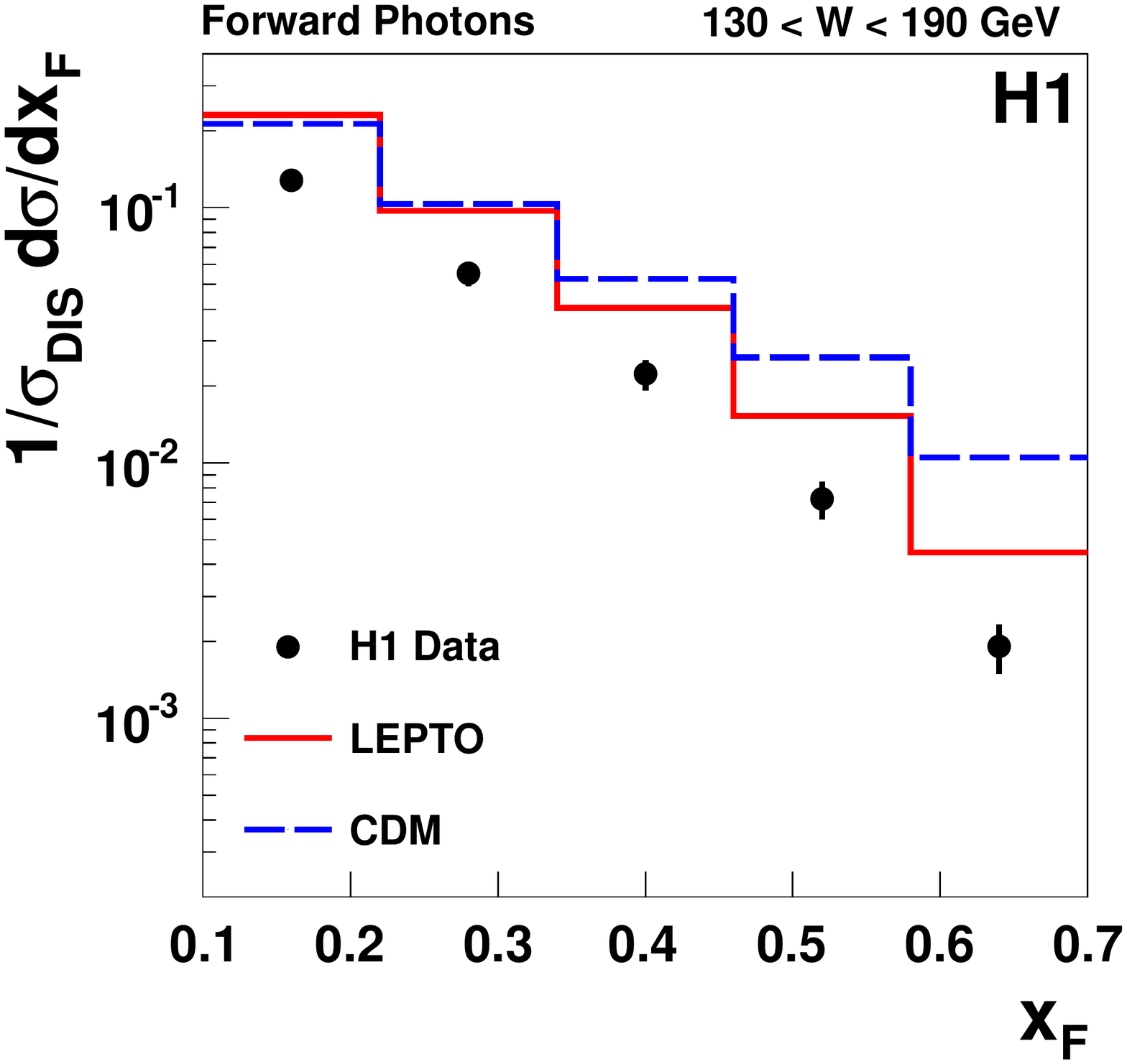}
}
\\[-35mm]
\hspace*{20mm}{\small (a)\hspace*{36mm}(b)}\\[16mm]
\mbox{\hspace*{-1mm}
\includegraphics[width=40mm]{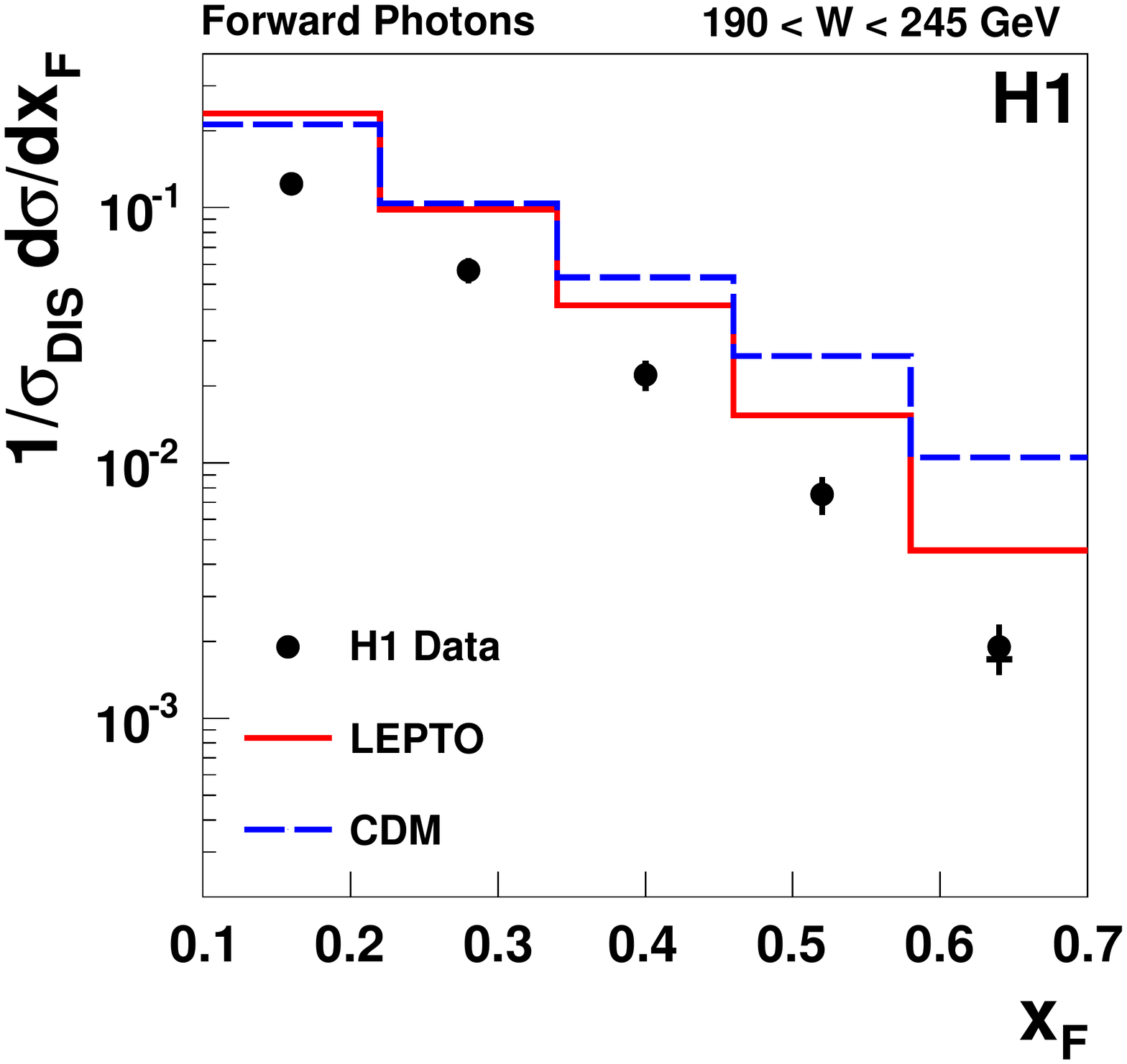}
\includegraphics[width=40mm]{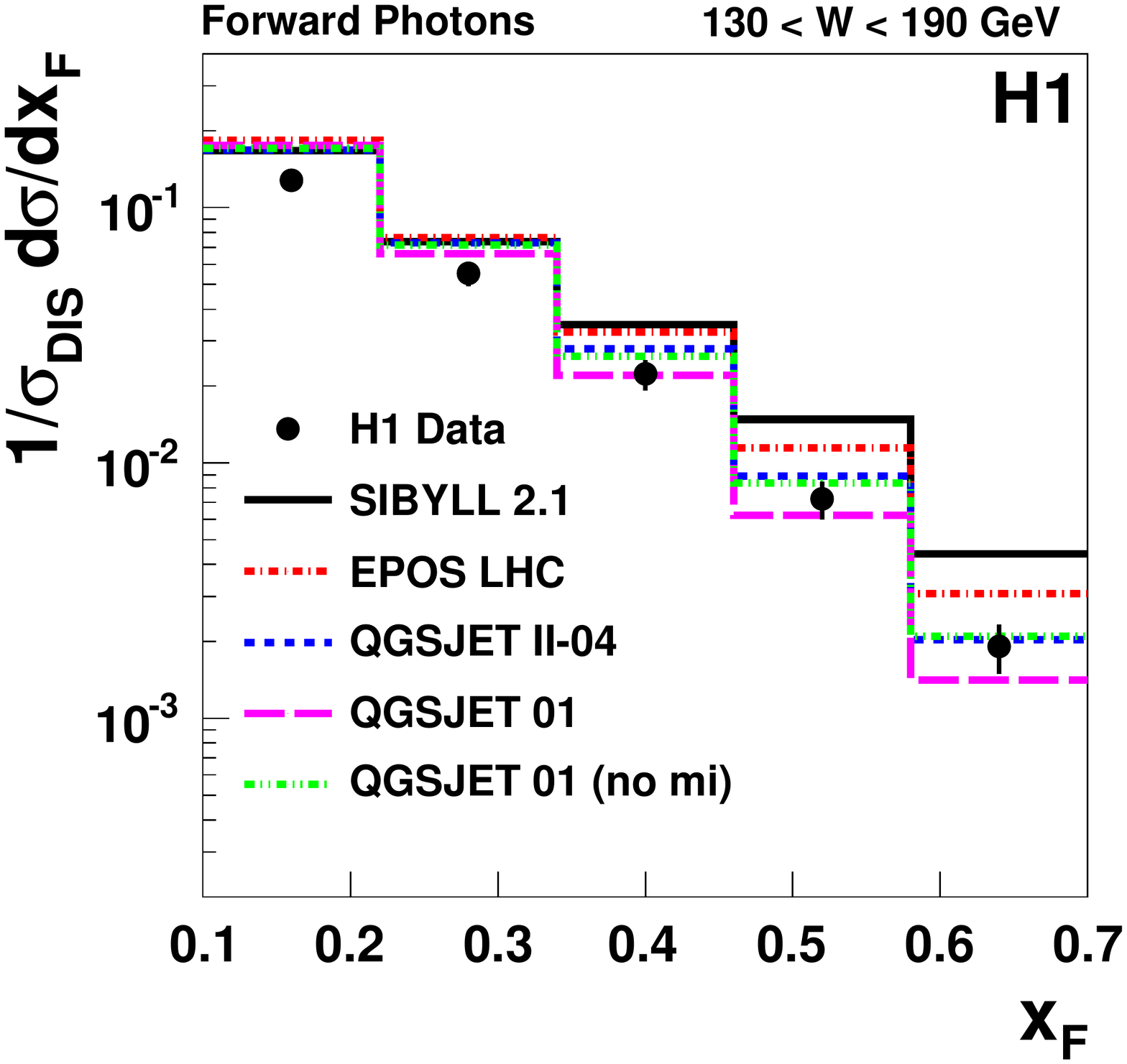}
}
\\[-35mm]
\hspace*{20mm}{\small (c)\hspace*{36mm}(d)}\\[26mm]
\end{centering}
\caption{$x_F$ distributions for forward photons in H1
for a series of different centre-or-mass ranges (a-c)
compared to LEPTO and CDM, and an
example (d) of comparisons to other models.}
\label{fig3}
\end{figure}

\begin{figure}[h!]
~\\[-12mm]
\begin{centering}
\mbox{\hspace*{-1mm}
\includegraphics[width=40mm]{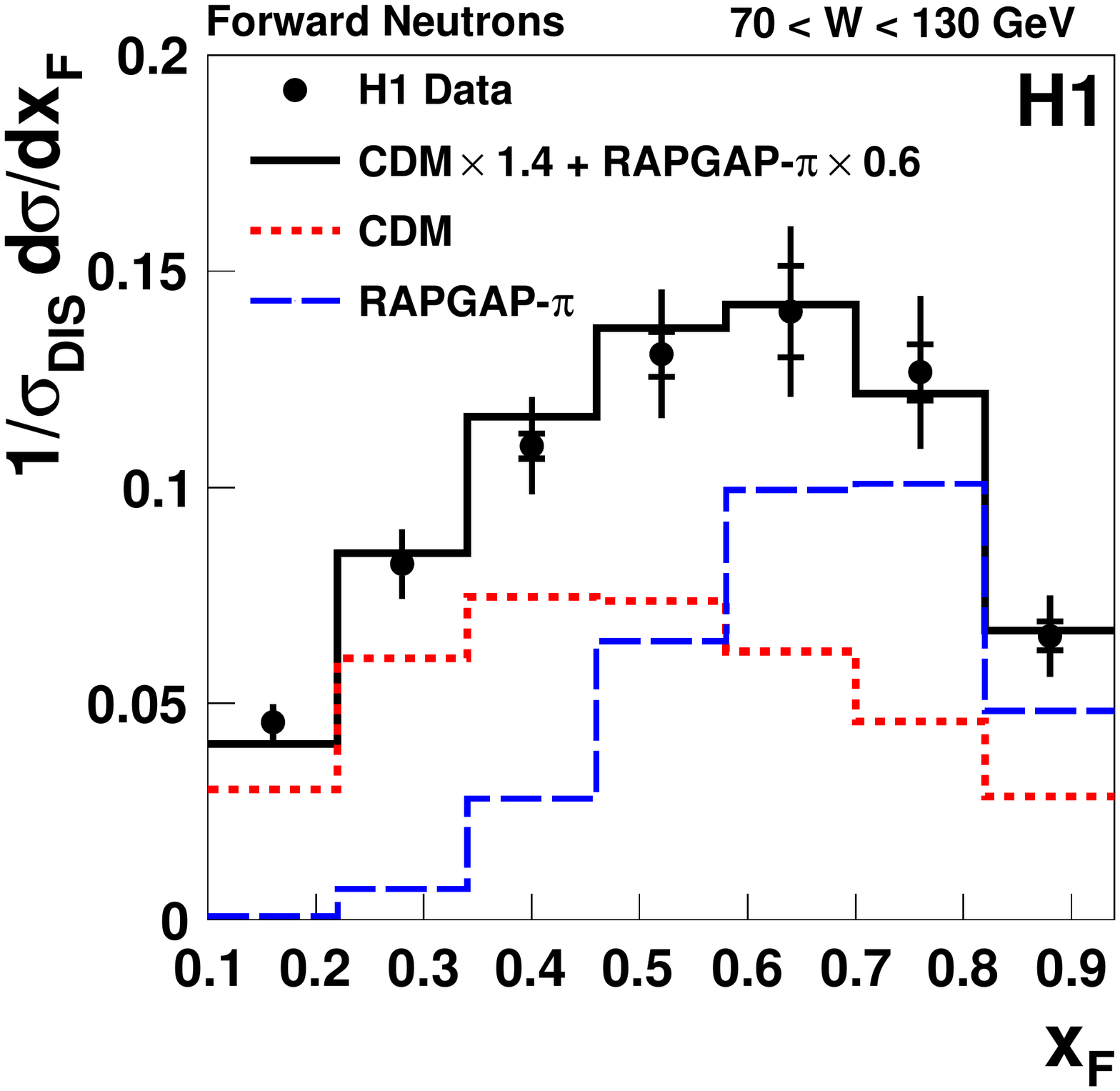}
\includegraphics[width=40mm]{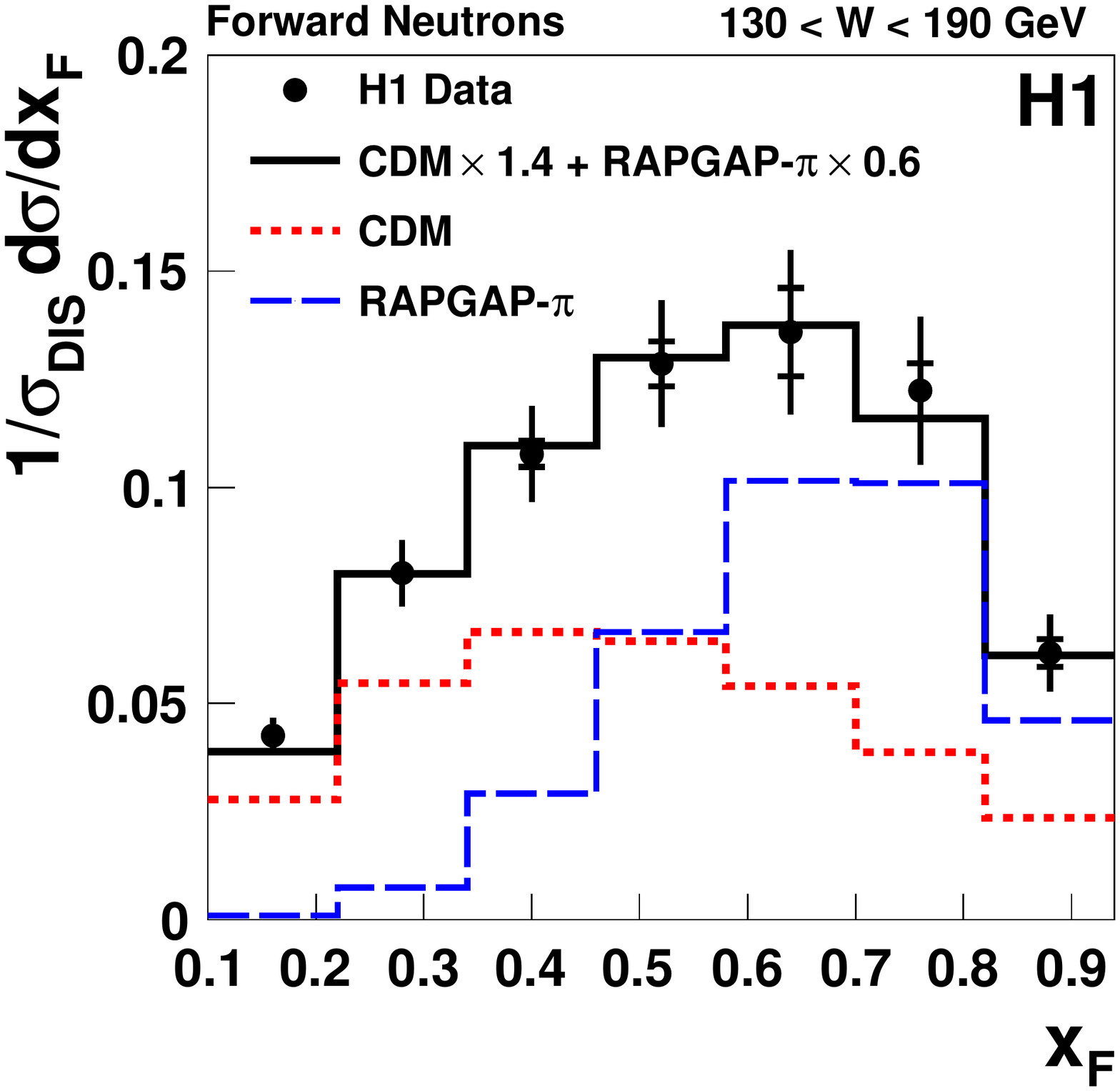}
}
\\[-36mm]
\hspace*{20mm}{\small (a)\hspace*{36mm}(b)}\\[16mm]
\mbox{\hspace*{-1mm}
\includegraphics[width=40mm]{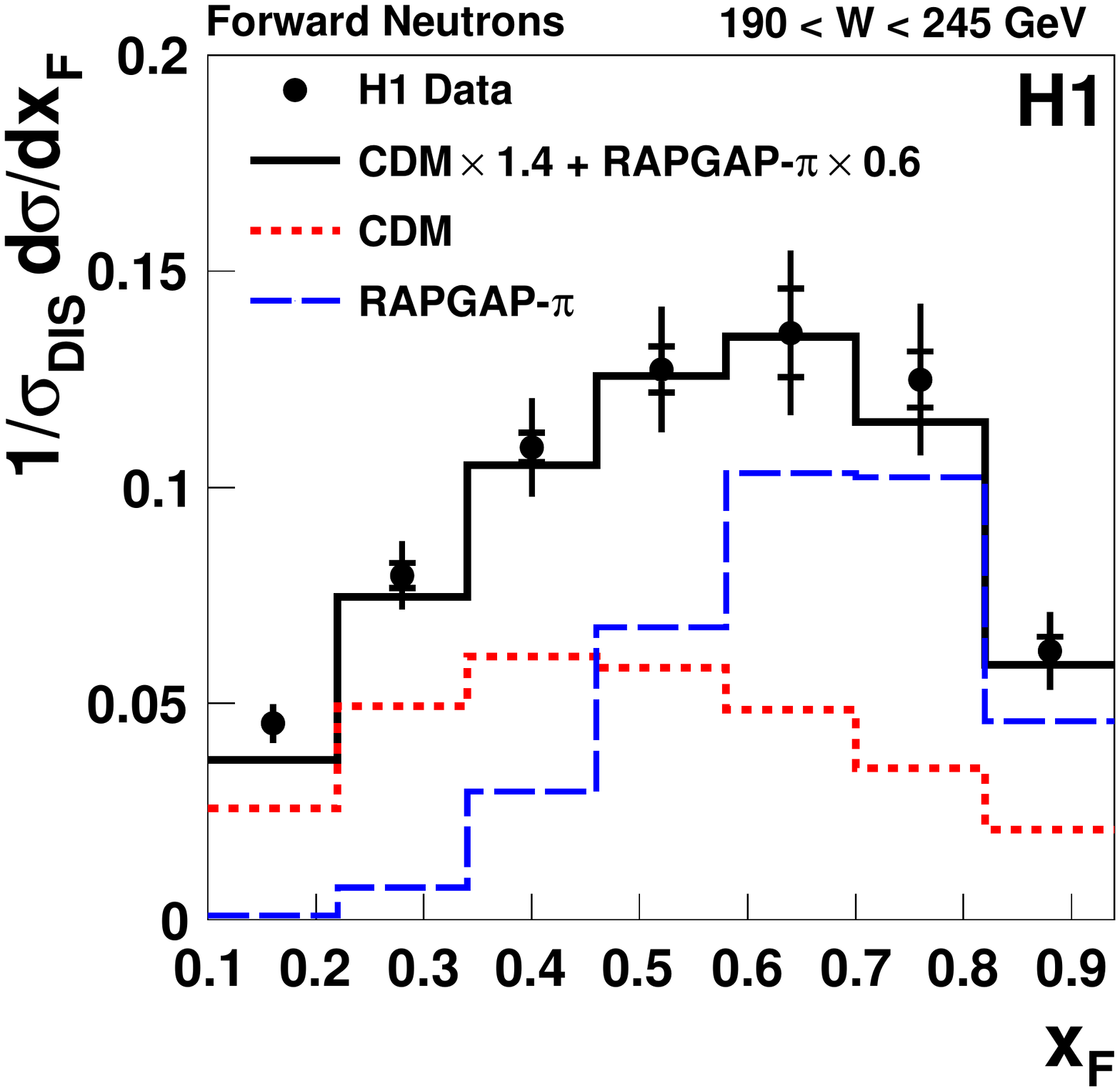}
\includegraphics[width=40mm]{d14-035f6c}
}
\\[-36mm]
\hspace*{20mm}{\small (c)\hspace*{36mm}(d)}\\[26mm]
\end{centering}
\caption{$x_F$ distributions for forward neutrons in H1
for a series of different centre-or-mass ranges (a-c)
compared to LEPTO and CDM, and an
example (d) of comparisons to other models.}
\label{fig4}
\end{figure}
\begin{figure}[h!]
~\\[-10mm]
\begin{centering}
\mbox{\hspace*{-1mm}
\includegraphics[width=40mm]{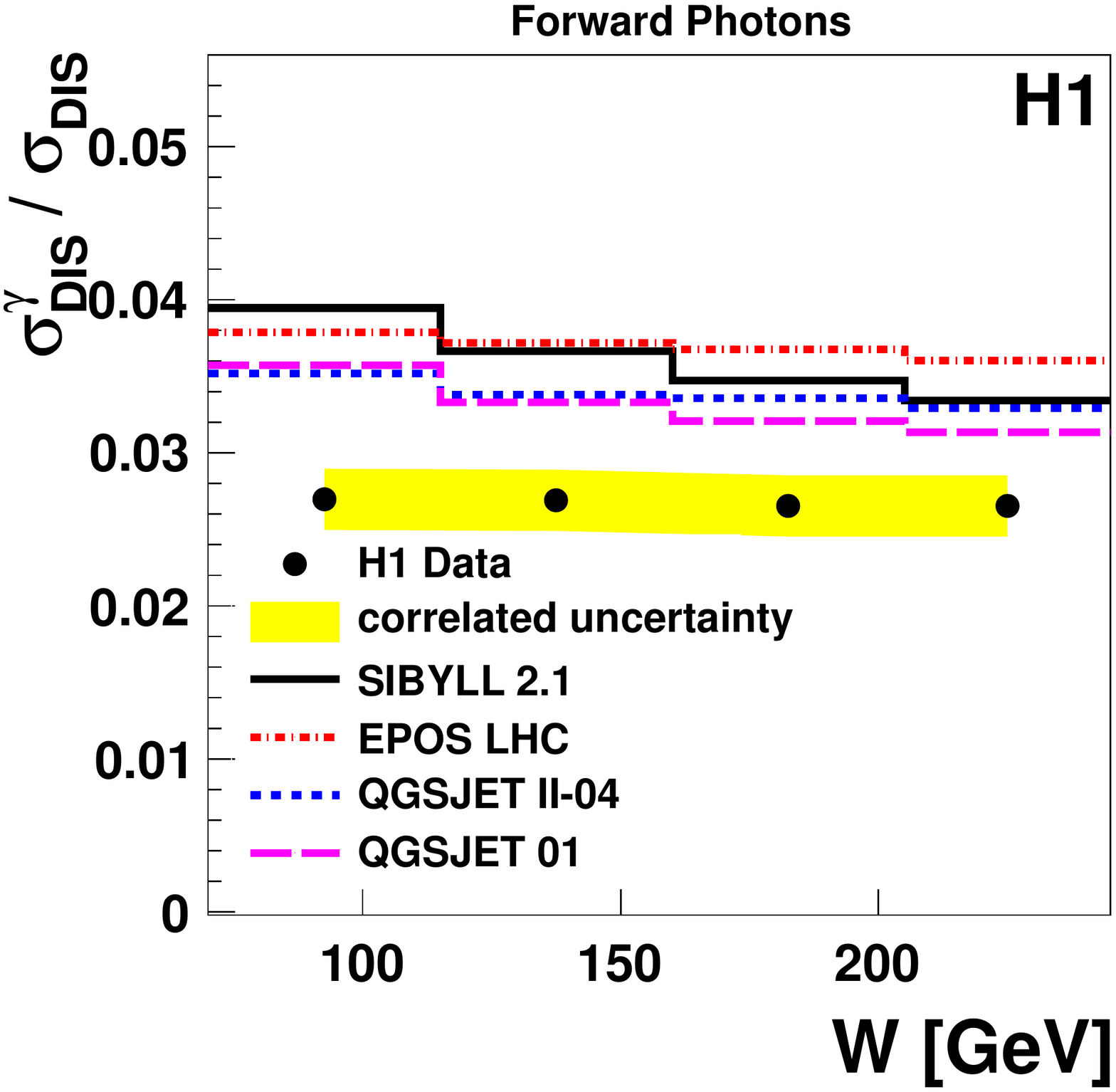}
\includegraphics[width=40mm]{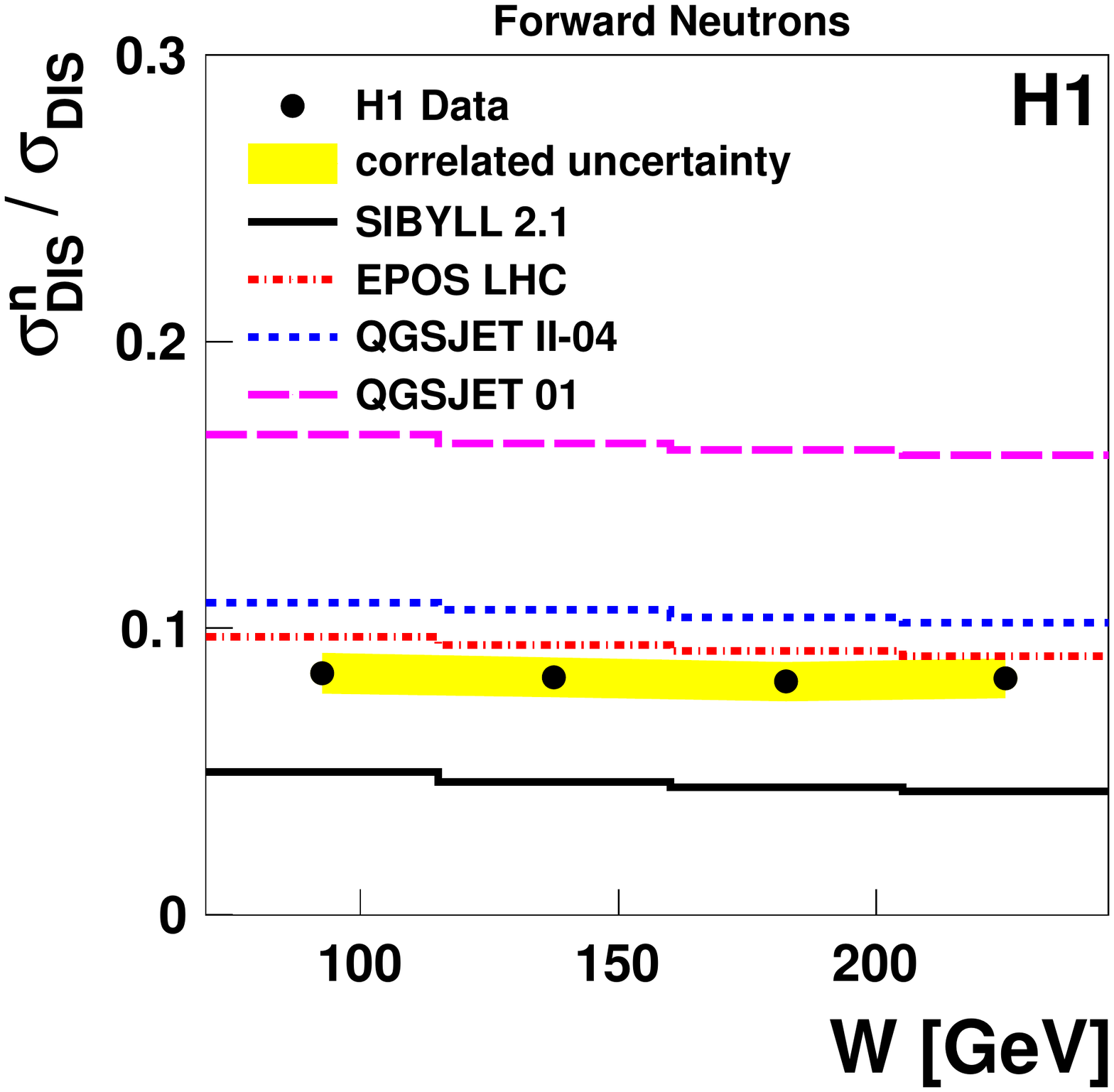}
}
\\[-36mm]
\hspace*{20mm}{\small (a)\hspace*{36mm}(b)}\\[26mm]
\end{centering}
\caption{Normalised cross sections as a function of $W$
for (a) forward photon production (b) forward neutron production,
compared to a number of theoretical predictions.}
\label{fig2}
\end{figure}

\section{Forward photons and neutrons}
Positioned 106 m downstream of the interaction beyond some bending
magnets, the H1 Forward Neutron Counter was able to detect and
distinguish between very forward emerging photons and neutrons. Both
types of particle emerged from the decay of excited proton states,
while the neutrons were also produced through colour singlet exchange
processes.  The production rates of these particles were measured by H1,
with a particular emphasis on establishing whether Feynman scaling holds, 
i.e.\ whether cross sections as a function of the centre-of-mass energy $W$
are independent of the Feynman variable $x_F = 2p_\parallel^*/W$.
The centre-of-mass energy $W$ is defined as $\sqrt{ys - Q^2}$ and all variables
are defined in the laboratory frame.

Measurements were made over the range $6<Q^2<100$ GeV$^2$.  Models
tested included LEPTO, based on Lund string fragmentation, and
RAPGAP combined with ARIADNE, which uses a colour dipole  (CDM)
formalism.  A further set of theoretical calculations made use of
models that were originally constructed to simulate cosmic ray
showers, but were adapted for the $ep$ context.  These were 
SIBYLL and QGSJET, which are reggeon-based and were interfaced
using PHOJET, and EPOS LHC, which is based on a parton
model, modifying the treatment of central diffraction according to LHC
measurements.

Figure \ref{fig3} presents cross sections for forward photons,
normalised relative to the inclusive DIS cross section, as a function of
$x_F$, for three different ranges of $W$.  It can be seen that the
cross sections show no discernable variation with $W$, confirming the
principle of Feynman scaling.  The colour dipole model does not fit
well the shape of the distribution.  The cosmic-ray based models show
a better ability to fit the data, but SIBYLL fails.  The
absolute values of the normalised cross sections are not reproduced
well by all the models.  Figure \ref{fig4} presents the corresponding
neutron cross sections.  Feynman scaling is again confirmed, and a
combination of RAPGAP and CDM effectively reproduces the shape
and the absolute value of the normalised cross sections.
Of the cosmic-ray based models, there is greater variability than with the
photons, and only EPOS LHC can be considered reasonably satisfactory.

Figure \ref{fig2} shows the total photon and neutron cross sections,
normalised to the total DIS cross sections, for photons and neutrons
respectively.  Comparison is made to the cosmic-ray based models,
showing that all are consistently high in the photon case, but vary
widely in the neutron case with EPOS LHC being best.  Both sets
of data distributions are flat in $W$.

\section{Isolated ``prompt'' photon production}
The ZEUS Collaboration have performed studies of isolated
high-energy photons, known as ``prompt'' photons, in photoproduction,
measuring the basic photon and jet variables~\cite{z1} and a number of
other kinematic quantities~\cite{z2}.  The photons were detected in
the electromagnetic section of the ZEUS Barrel Calorimeter, and jets
were identified by ``energy-flow objects'' constructed from
calorimeter energy deposits and measured tracks.  The photon signal
was accompanied by a background arising from neutral hadrons such as
$\pi^0$ and $\eta$ mesons, which characteristically gave broader
clusters of firing cells in the calorimeter. For each measured bin, a
fit was performed to the width of the calorimeter cell cluster, so as
to extract the photon signal.

\begin{figure}[t!]
~\\[-35mm]
\begin{centering}
\hspace*{9mm}
\includegraphics[width=60mm]{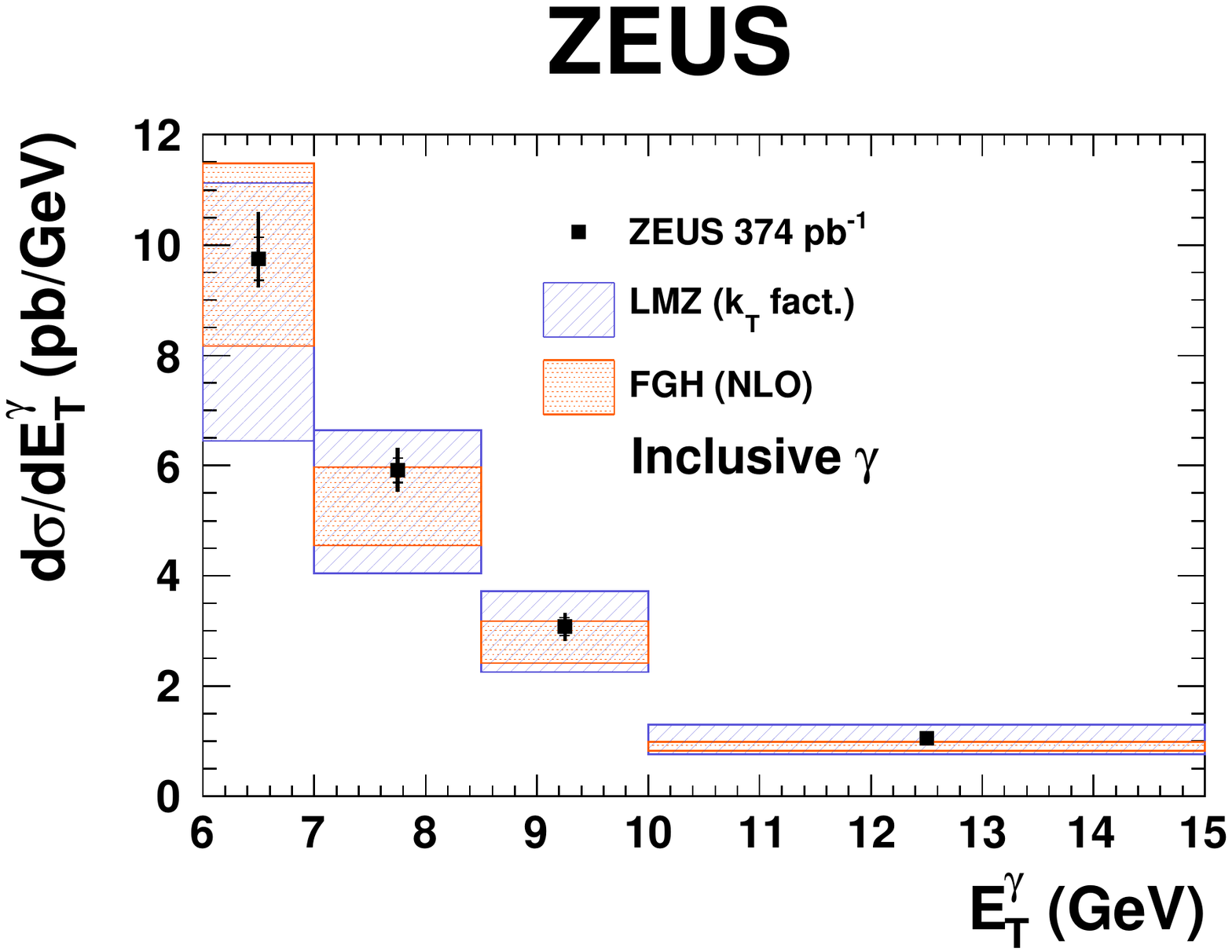}
\\[-34mm]\hspace*{39mm}(a)\\[-10mm]
\hspace*{9mm}
\includegraphics[width=60mm]{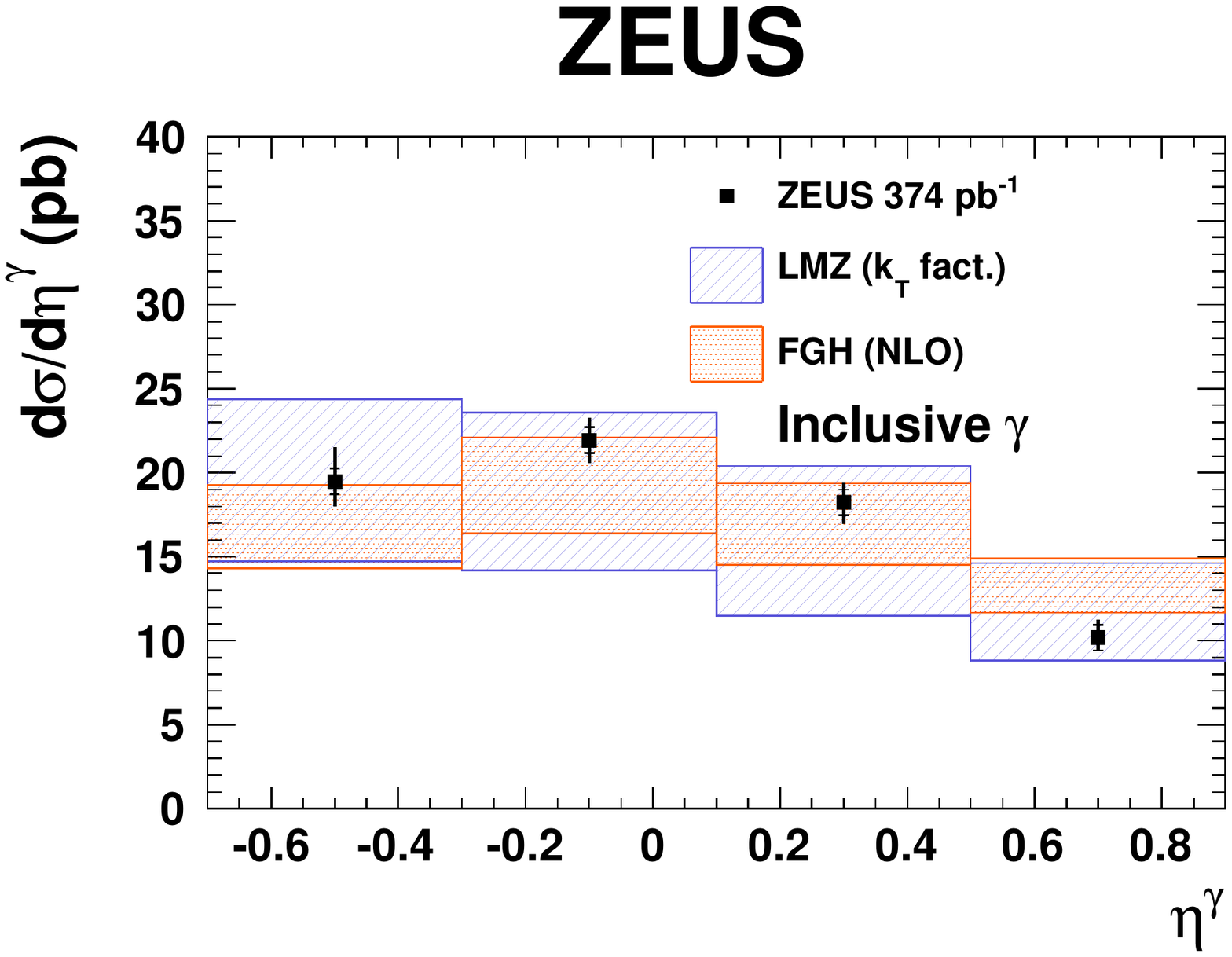}
\\[-34mm]\hspace*{39mm}(b)\\[24mm]
\end{centering}
\caption{Distributions in (a) transverse energy 
and (b) pseudorapidity
of inclusively
produced isolated photons in ZEUS.}
\label{fig5}
\end{figure}

\begin{figure}[t]
~\\[-48mm]
\begin{centering}
\hspace*{1mm}
\includegraphics[width=78mm]{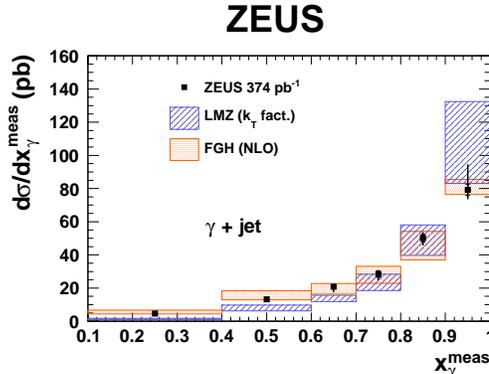}
\\[-34mm]\hspace*{15mm}\\[23mm]
\end{centering}
\caption{Distributions in $x_\gamma^\mathrm{meas}$
for the prompt-photon + jet final state in
photoproduction.
}
\label{fig5a}
\end{figure}

\begin{figure}[t]
~\\[-48mm]
\begin{centering}
\hspace*{0mm}
\includegraphics[width=85mm]{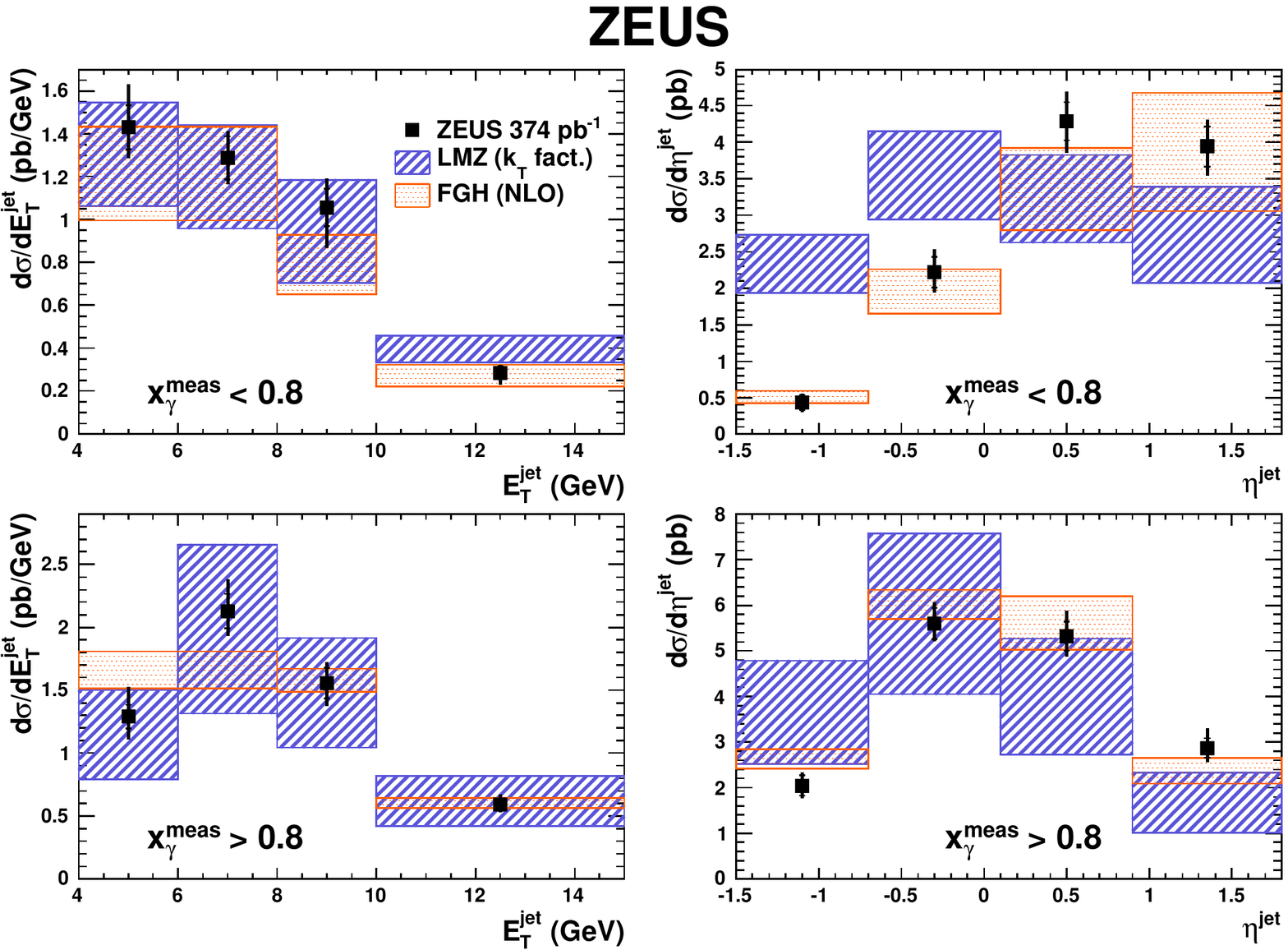}
\\[-34mm]\hspace*{15mm}\\[28mm]
\end{centering}
\caption{Distributions in jet transverse energy
and pseudorapidity for $x_\gamma^\mathrm{meas}$ values less
than and greater than 0.8}
\label{fig6}
\end{figure}

Two theoretical models were tested.  That of Fontannaz, Guillet and
Heinrich (FGH) consisted of a standard next-to-leading-order QCD
calculation augmented by a box diagram contribution and a jet
fragmentation contribution.  A second model, by Lipatov, Malyshev and
Zotov (LMZ) used unintegrated parton distributions and an
initial-state parton cascade.

Measurements were made for 
photon and jet transverse energies above 6 GeV and 4 GeV respectively.
An important phenomenological quantity is $x_\gamma^\mathrm{meas}$,
defined as the fraction of the final-state $E - p_Z$ that is contained in
the photon and the jet, hence giving a relativistically-invariant measure of the fraction of the
incoming photon energy that takes part in the QCD scattering process.
In direct processes, all the photon energy takes part in the QCD scatter,
while in resolved processes the photon acts as a source of partons.
When measured, smearing due to fragmentation and higher-order processes
becomes introduced. Cross sections were evaluated for the entire
$x_\gamma^\mathrm{meas}$ range, and for ranges below and above a value of
0.8, which denoted resolved-enhanced and direct-enhanced regions of the kinematics.
An isolation criterion is imposed, such that a photon must contain at least 90\%
of the energy of the jet-like object (which may be just the photon itself) that
contains it.  This reduces backgrounds and the effects of the fragmentation component,
which is difficult to model accurately.

Figure \ref{fig5} presents cross sections for inclusive photons as
functions of transverse energy and of pseudorapidity.  Both the
theoretical models give a good description of the distributions.
The cross section in  $x_\gamma^\mathrm{meas}$ is shown in fig.~\ref{fig5a},
where the enhancement towards the value of unity is due to the direct
photoproduction process, while the resolved process gives a broader distribution.
The theoretical FGH is in good agreement with the data, the LMZ model slightly less
so.  

For the second set of new  measurements~\cite{z2} an updated
version of the LMZ calculation was used.  The corresponding
distributions when the photon is accompanied by a jet show good
agreement with both models.  Distributions for the transverse energy
of the jet are given for the two
$x_\gamma^\mathrm{meas}$ ranges in fig.~\ref{fig6} and show that
while the direct-enhanced region is well-described by both models, the
resolved-enhanced region has its pseudorapidity distribution poorly
described by the LMZ model, possibly indicating a defect in the
modelling of the initial-state parton cascade.

\begin{figure}[t!]
~\\[-10mm]
\begin{centering}
\hspace*{7mm}
\includegraphics[width=95mm]{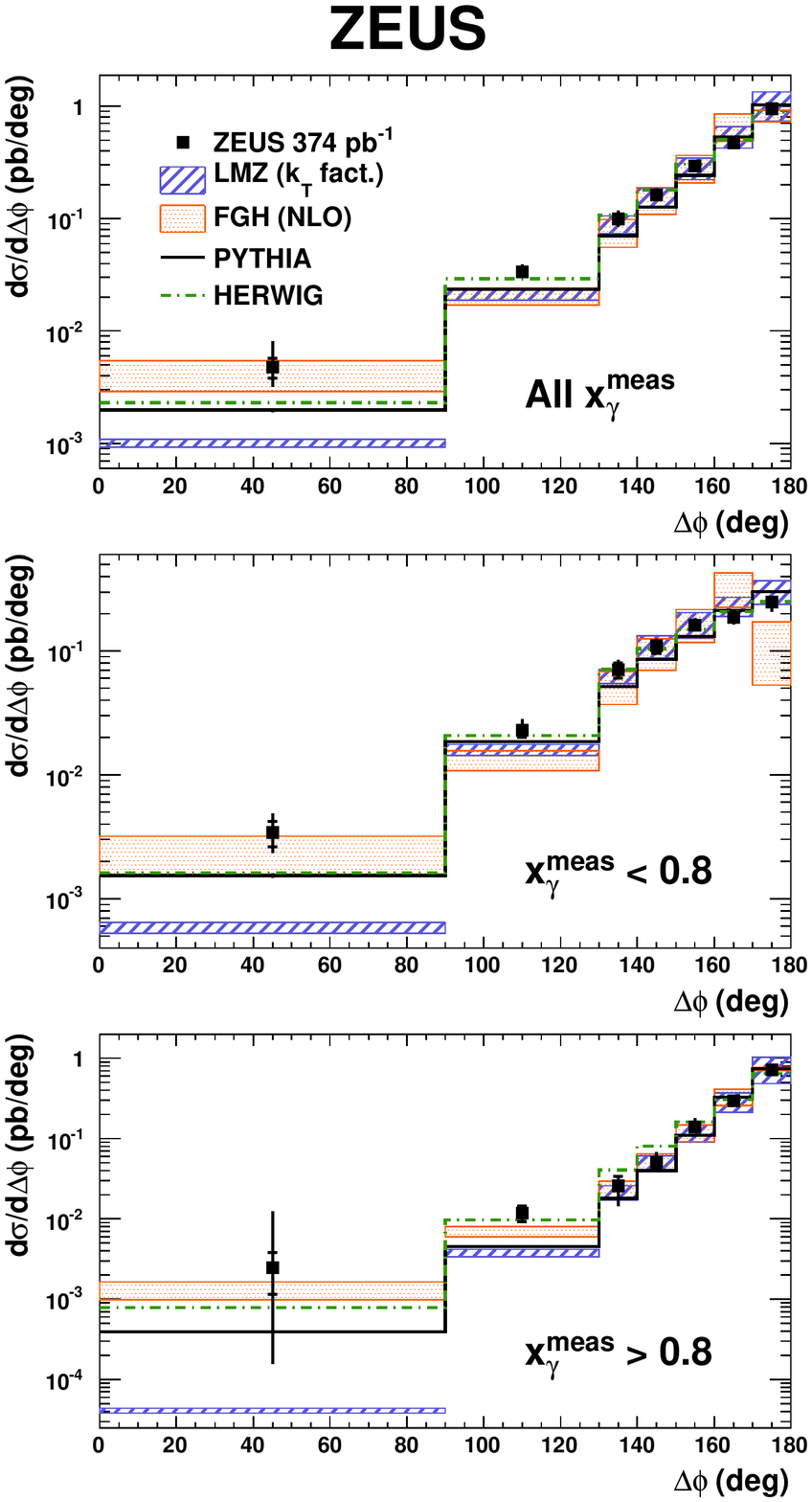}
\end{centering}
\caption{Distributions in azimuthal distance
between photon and jet for the entire  $x_\gamma^\mathrm{meas}$ range 
and for values less
than and greater than 0.8, compared to calculations.}
\label{fig7}
\end{figure}

Figure \ref{fig7} shows that the difference between the azimuths of the photon and
the jet is well-described both by the parton-level models already mentioned, and also
by the parton-shower Monte Carlos PYTHIA and HERWIG.  This suggests that
the showering mechanisms used in the latter are in some circumstances a good 
representation of a higher-order parton calculation.

\section{Conclusions}
The HERA experiments have continued to produce new and innovative measurements 
of hadronic final states, capable of testing state-of-the art theoretical calculations.
Further results are expected over the coming year.




\nocite{*} \bibliographystyle{elsarticle-num} \bibliography{martin}



\vfill

\end{document}